# In-resonator variation of waveguide cross-sections for dispersion control of aluminum nitride micro-rings

**Hojoong Jung, Menno Poot, and Hong X. Tang**[*]

*Department of Electrical Engineering, Yale University, New Haven, Connecticut 06511, USA*
[*]*hong.tang@yale.edu*

**Abstract:** We propose and demonstrate a dispersion control technique by combination of different waveguide cross sections in an aluminum nitride micro-ring resonator. Narrow and wide waveguides with normal and anomalous dispersion, respectively, are linked with tapering waveguides and enclosed in a ring resonator to produce a total dispersion near zero. The mode-coupling in multimoded waveguides is also effectively suppressed. This technique provides new degrees of freedom and enhanced flexibility in engineering the dispersion of microcomb resonators.

## 1. Introduction

Optical frequency combs [1,2] are of considerable interests for applications including optical clocks, frequency metrology and molecular spectroscopy [3-5]. In particular, frequency comb generation from micro-resonators using four wave mixing (FWM) has recently been explored for its simplicity and compactness with various materials platforms, such as silica toroids [6], doped silica [7], silicon [8], silicon nitride (SiN) [9], AlN [10], diamond [11], crystalline $CaF_2$ [12] and $MgF_2$ resonators [13]. The large enhancement of circulating optical power in these high Q resonators has led to dramatic reduction of threshold of comb generation.

Dispersion control of micro-resonators is critical for Kerr-comb generation as it limits the comb bandwidth [2]. In optical waveguides, dispersion can be controlled simply by adjusting the waveguide width and height [8-11]. Other methods have been studied, for example, by changing the sidewall angle of silica toroids [14], oxidation of silicon micro-disks [15], atomic layer deposition on SiN [16], microstructuring of $MgF_2$ resonators [17], etc. These dispersion engineering techniques, however only uniformly adjust the waveguide cross section of the whole resonator that does not have many degree of freedoms for complete dispersion engineering. Here, we propose that the dispersion can be compensated by combining two waveguides of different cross sections with opposite signs of dispersion within a single ring resonator. This method also can eliminate the mode crossing in multimode waveguide, which is inevitable for anomalous dispersion within the current design. We apply this design concept to AlN micro-resonators and experimentally realized a near-zero dispersion without mode crossing.

## 2. Experiment

We previously demonstrated near zero dispersion with 650 nm-thick and 3.5 μm-wide AlN waveguides [10]. The goal here is, however, to more generally control the dispersion by combining waveguide segments with different dispersion properties. Figure 1 shows the schematics of our device. In a single waveguide resonator, two sections with different widths and dispersions are made. In a narrow waveguide, which has normal dispersion (i.e. negative group velocity dispersion, GVD < 0), the longer wavelength light (red curve) propagates faster than the shorter wavelength light (blue curve). This dispersion is compensated as they propagate in the wide waveguide region which has anomalous dispersion (GVD > 0). As a result, a total dispersion of zero at the desired wavelength can be obtained. The different waveguide segments are linked by adiabatic tapers to minimize optical loss. This structure has two major advantages: 1) One section of waveguide in the resonator is chosen to be single-mode, so that multi-mode resonances or mode crossings are soothed. 2) More flexible dispersion control is possible. As an example, a dispersion curve crossing zero at 775 nm and 1550 nm will be discussed.

We fabricate this device from a 700 nm-thick AlN film. 140 nm PECVD $SiO_2$ is first deposited over the whole film as a hard mask to enhance the etching selectivity. 300 nm-thick MaN resist is used for patterning, and a reflow technique [9] is used to decrease the sidewall roughness of the waveguides. Two step reactive-ion etching (RIE) is performed, one for etching the $SiO_2$ mask and another for etching AlN. Finally the device is covered by a 3 μm PECVD $SiO_2$ top cladding and polished for fiber butt coupling. Due to the reflow the MaN resist becomes trapezoidal and the fabricated waveguides have a 19 - 22$^0$ sidewall angle and the top width is a slightly narrower than nominal value. The designed resonator has a total circumference of 1.87 mm which corresponds to a 70 – 80 GHz free spectral range (FSR) depending on the group index of the mode.

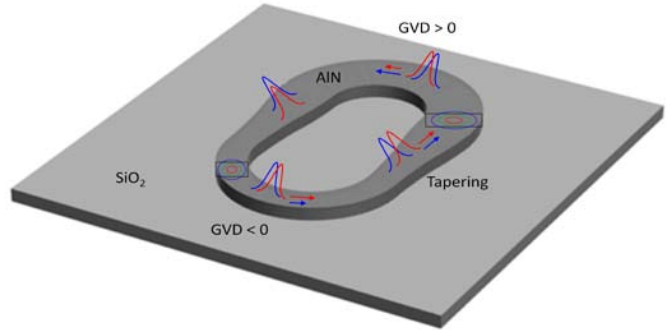

Fig. 1. A schematic of a multi-segment resonator that employs waveguides of different cross-sections for dispersion compensation (not to scale). The red and blue curves are the light of longer and shorter wavelengths, respectively. The length of the arrows represent the speed of light for each wavelength in different waveguide sections. Mode profiles in narrow and wide waveguides are also sketched.

A micrograph of a fabricated device is shown in Fig. 2(a). The bending radii are 50 μm which are large enough to give, based on our simulations, ignorable (- 5 ps / nm · km) effects on the dispersion. To measure the exact FSR, a continuous wave (CW) step laser and a wavemeter are used. The polarization is adjusted to the TE-like mode using a fiber polarization controller (FPC), and the optical power in the input waveguide is estimated to be 100 μW. The measured transmission spectrum is shown in Fig. 2(b). The cross section of this device has a height of 700 nm and top width of 350 nm. The intrinsic quality factor ($Q$) of this resonator is 200,000, and the corresponding waveguide loss is 1.9 dB/cm [18]. The FSRs near wavelength λ = 1551 nm and 1578 nm are 75.31 GHz and 75.77 GHz, respectively as shown in Fig. 2(c). This variation of the FSRs with wavelength is extracted to estimate the dispersion of our resonators.

The measured FSR and dispersion of a waveguide with top width of 350 nm are shown in Fig. 3(a), display a strong anomalous dispersion of -1070 ± 80 ps / nm · km. Due to the large bottom width of the waveguide (850 nm), the waveguide also supports the second order TE mode, and five mode crossings are observed in this wavelength range. We then fabricate wider waveguide with top width of 700 nm and measure a very small dispersion, -13 ± 40 ps / nm · km [Fig. 3.(b)]. Note that the FSR data points near mode coupling range are not considered in estimating the dispersion. This device has an intrinsic quality factor of 250,000 and optical loss of 1.5 dB/cm. The insets in Fig. 3(a) and (b) show the scanning electron microscope (SEM) images of their cross sections and corresponding $TE_{00}$ mode power profile. Now we know two different waveguides with dispersions. When light propagates through the multi-segment resonator, the effective dispersion, $\bar{D}$, is a linear summation of dispersions in each segment :

$$\bar{D}(\lambda) = \frac{1}{C} \oint_{resonator} D(\lambda, s) ds \qquad (1)$$

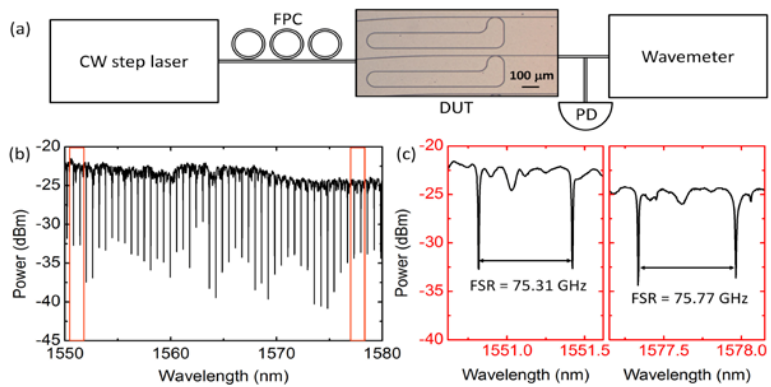

Fig. 2. (a) The measurement setup. CW, continuous wave; FPC, fiber polarization controller; DUT, device under test; PD, photo diode. (b) A portion of the transmission spectrum measured using the wavemeter. (c) Zoom of the spectrum near 1551 nm and 1578 nm wavelength indicating different FSRs due to uncompensated dispersion.

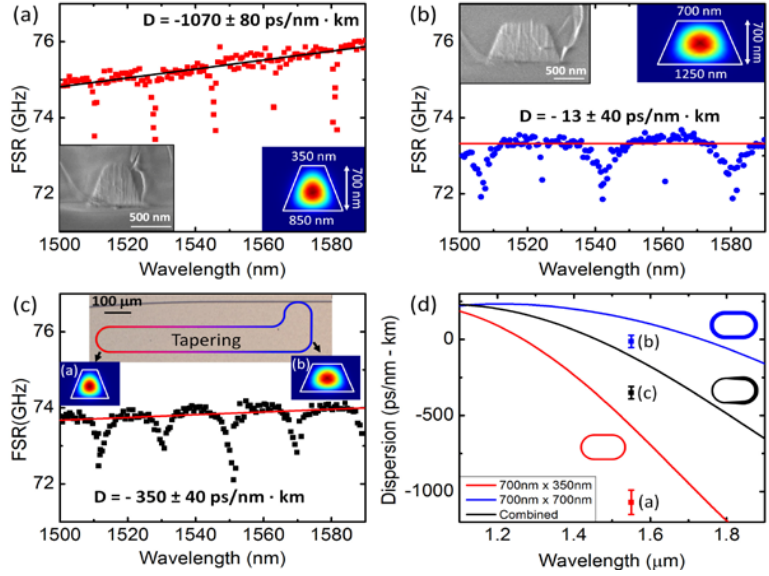

Fig. 3. The measured FSR and dispersion of AlN resonators of varying cross-sectional configurations fabricated from 700 nm-thick thin films. (a) FSR measured from a ring resonator with a uniform, trapezoidal cross-section. The left inset is an SEM image of the cross section, and the right inset is the corresponding mode power profile. (b) FSR taken from a resonator with wider cross section (top width: 700nm). (c) FSR measured in a ring resonator containing waveguides with two cross sections. Inset shows the layout of the micro-resonator: the left (red) part of the resonator has a 350 nm top width, and they are tapered up to the right (blue) part which has a 700 nm top width. (d) The simulated dispersion curve and measured dispersion value for single-width (red and blue) and multi-width (black) resonators. The length of each single-width waveguide and tapering between them are considered through Eq. (1).

, where $C$ is the total length (i.e., circumference) of the resonator; $D(\lambda,s)$ is the dispersion at wavelength $\lambda$, and segment at position $s$.

To test this idea, we fabricated resonators which consist of the two cross sections. The inset of Fig. 3(c) shows the resonator with two cross sections and tapering regions between them. The left (red) part of resonator has a 350 nm top width (which is the same as the device indicated in Fig. 3(a)), and right (blue) part of resonator has a 700 nm top width (as in Fig. 3(b)). The center part is the adiabatic tapering region to link the two different widths with minimal loss. The FSR is measured and from its wavelength dependence, $\bar{D}$ = -350 ± 40 ps / nm · km dispersion is extracted [Fig. 3(c)]. The net dispersion thus lies in between the values of the two single-width resonators, but it is closer to that of the wider waveguide which is longer than the narrow waveguide (0.514 mm vs 0.157 mm) in this resonator. The tapering region (600 μm long, tapering angles 0.02°) should be adiabatic to prevent quality factor degradation. [19-21]. The intrinsic quality factor is 200,000, which is the same as the narrow-width-waveguide-only resonator. The corresponding optical loss is 0.35 dB/round-trip.

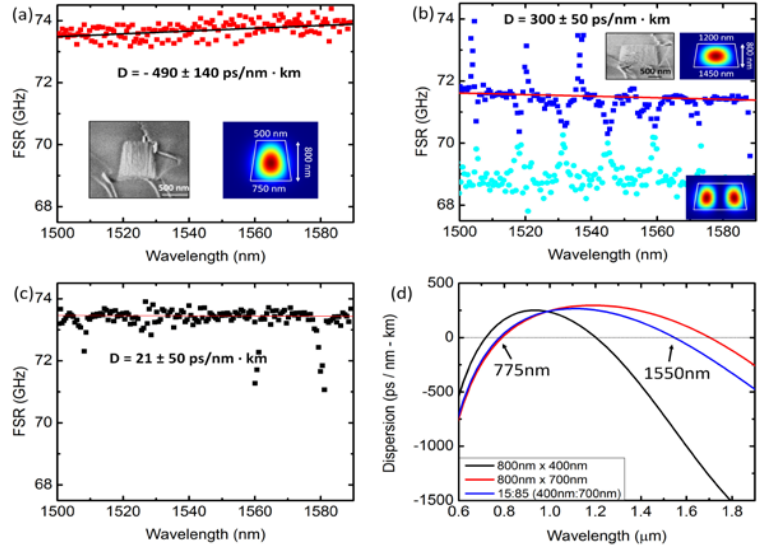

Fig. 4. Measured FSR and dispersion for waveguides fabricated from 800 nm-thick AlN films with their SEM images and mode profiles. (a) The measured FSR from a resonator of 500 nm-wide at the top of the trapezoid has a normal dispersion. (b) Anomalous dispersion measured from a waveguide of 1200 nm-width. Blue points are for the fundamental TE mode, and the cyan points are for the second order TE mode. (c) The near-zero dispersion is obtained by combining waveguides with two cross sections, 800 nm × 500 nm and 800 nm × 1200 nm. (d) Simulated dispersion curves showing simultaneous zeroing of the dispersion at 775 nm and 1550 nm wavelengths (blue) by combining 800 nm × 400 nm (black) and 800 nm × 700 nm waveguides (red).

Considering the propagation losses from waveguides of different widths, the estimated loss from the tapering structure is extracted to be 0.02 dB/each. This value is much smaller than the waveguide propagation loss, which means the $Q$ of resonators is not limited by the tapering region.

Figure 3 (d) compares the simulated dispersion curves and experimental data points of single-width and multiple-width resonators. Here, the blue line is the simulated dispersion for the 700 nm-wide waveguide. The red line is for the 350 nm-wide waveguide and the black line is the length weighted dispersion combining all the waveguide segments including the tapers as calculated using Eq. (1). The three red, blue, and black data points with the error bar are the dispersion obtained from Fig. 3(a) to (c), respectively. The offset between the simulation and experimental results could be due to a deviation of the material dispersion used in the simulation from the true material dispersion. Furthermore, uncertainties in the dimension measurement from the SEM (~20 nm) may also contribute to the mismatch. Nevertheless, the black point, which is for the combined resonator, falls in between the blue and red data points. The dispersion of the black point can be adjusted by changing the length ratio of the wide and narrow waveguide. If the length of wide waveguide (blue) is increased, the combined dispersion can go to anomalous dispersion region, and vice versa.

After having verified the concept using 700 nm-thick AlN films, we fabricate 800 nm-thick devices which can, based on the simulation, provide higher anomalous dispersion than 700 nm waveguides. For 800 nm-thick AlN, we use 500 nm-thick HSQ resist (FOx-16, Dow Corning) for better width control in patterning. For the normal dispersion waveguide, a top width of 500 nm and height of 800 nm waveguide is used. The FSR and dispersion in this device are 73.5 – 74 GHz and -490 ± 140 ps / nm · km respectively as shown in Fig. 4(a). This graph shows no mode crossings due to the narrow waveguide structure. Based on the simulation, the waveguide has a weakly confined second order mode, but it is not observed experimentally due to the leakage from bending and surface scattering. The inset of Fig. 4(a) shows the SEM image and the $TE_{00}$ mode power profile of the waveguide which has a $9^0$ sidewall angle. For the anomalous dispersion waveguide, a width of 1200 nm is used. In this case, the TE$_{00}$ mode of the waveguide shows a strong anomalous dispersion (300 ± 50 ps / nm · km, Fig. 4(b)). Using these two waveguides with normal and anomalous dispersion, near zero dispersion resonators can be obtained: The dispersion of the combined resonator is 21 ± 50 ps / nm · km which is very close to zero [Fig. 4(c)]. Although a strong mode coupling between the TE$_{00}$ (blue points) and TE$_{01}$ (cyan points) modes was observed in the 1200 nm-wide, multimode waveguide [Fig. 4(b)] [22], it is strongly suppressed in the combined resonator [Fig. 4(c)] due to the inclusion of the narrow waveguide. Some mode crossings near 1560 nm and 1580 nm are still visible due to residual light survived in the TE$_{01}$ mode despite of its weak guiding. By increasing the length of the resonator or by sharpening the bends, this could be suppressed even further. The intrinsic quality factors of the resonators used in Fig. 4(a) to (c) are 130,000, 180,000 and 120,000, respectively, again indicating that the segments can be combined without significant Q degradation. The combinational method is the only way to make a zero dispersion resonator without mode coupling so far.

In addition, our dispersion engineering technique is highly versatile. We can simultaneously tune the dispersion of two preselected wavelengths to zero by picking proper waveguide segments. If the tapering region is not considered, the average dispersion at wavelengths $\lambda_i$, can be expressed into a sum over all the different segments:

$$\bar{D}(\lambda_i) = \frac{L_1 D_1(\lambda_i) + L_2 D_2(\lambda_i) + L_3 D_3(\lambda_i) + \cdots}{C} \quad (2)$$

where $C$ is the total length of the resonator; $L_j$ is the length of the j$^{th}$ waveguide (j = 1..N), and $D_j(\lambda_i)$ is the dispersion at wavelength $\lambda_i$. In this equation, we can adjust $L_j$ and $D_j$ while keeping $C$ constant, to modify the average dispersions $\bar{D}(\lambda_i)$. For example, we consider two very disparate target wavelengths of 775 nm and 1550 nm; achieving zero dispersion at double wavelengths simultaneously is very important for comb generation in materials with second-harmonic generation. To satisfy the conditions $\bar{D}(775nm) = \bar{D}(1550nm) = 0$, we use two cross sections (N=2) of the same height (800 nm) but different widths (400 nm and 700 nm). Inserting this into Eq. (2) gives a linear system of equations that is readily solved; it is found that the length of the narrow and the wide waveguides should be 15 % and 85 % of the total waveguide length ($C$), respectively. The simulation result with these lengths is shown in Fig. 4(d). The black curve is the dispersion of narrow waveguide, and the red curve is for the wide waveguide. The blue curve is the combination of those waveguides and indeed gives zero dispersion at 775 nm and 1550 nm wavelength. The tapering sections are not considered in this simulation, but could easily be incorporated without loss of functionality.

## 3. Conclusion

In summary, we demonstrate a versatile dispersion-control technique by combining waveguides with different cross sections in a single resonator. The resonator with two different widths has a dispersion between those of its two single-width constituents. Our measurement results match well to the simulations. This technique can be used for dispersion engineering to enhance four-wave mixing and frequency comb generation. Also, this technique can be applied to broaden phase matching windows in nonlinear optical processes, such as second harmonic generation. If more cross sections are used, almost any desired dispersion curve can be constructed.

**Acknowledgments**

This work was supported by a Defense Advanced Research Projects Agency (DARPA). H.X.T. acknowledges support from a Packard Fellowship in Science and Engineering. Facilities used were supported by Yale Institute for Nanoscience and Quantum Engineering and NSF MRSEC DMR 1119826. The authors thank Michael Power and Dr. Michael Rooks for assistance in device fabrication.